\documentclass[structabstract]{aa}  

\usepackage{graphicx}
\usepackage{natbib}
\bibpunct{(}{)}{;}{a}{}{,}
\usepackage{txfonts}
\usepackage{lscape}
\usepackage{enumerate}
\usepackage{color}

\newcommand{\ka}{the Fe K$\alpha$ }
\newcommand{\eqw}{equivalent width }

\newcommand\T{\rule{0pt}{2.6ex}}
\newcommand\B{\rule[-1.2ex]{0pt}{0pt}}

\begin{document}
   \title{Properties of the integrated spectrum of serendipitous 2XMM catalogue sources}

 	\author{P. Chaudhary
           \inst{1}
           \and
           M. Brusa\inst{1}
 	  \and	
	  G. Hasinger\inst{2}
 	  \and
 	A. Merloni\inst{1,3}
	   \and
	A. Comastri\inst{4}
	 }

 \institute{Max-Planck-Institute for extraterrestrial Physik,
  Giessenbachstrasse, 1, D-85748, Garching bei M\"unchen, Germany\\
        \email{chaudhary@mpe.mpg.de}
        \and
	Max-Planck-Institute for Plasma Physics, Boltzmannstrasse 2, D-85748, Garching, Germany	
	\and
 	    Excellence Cluster Universe,
	  Boltzmannstrasse 2, D-85748, Garching bei Muenchen, Germany
	\and
	 INAF - Osservatorio Astronomico di Bologna, via Ranzani 1, 40127 Bologna, Italy \\
              }

\date{Received ; accepted}

 \abstract{}{}{}{}{} 
\abstract
{}
{Our analysis is aimed at characterizing the properties of the integrated spectrum of active galactic nuclei (AGNs) such as the ubiquity of the Fe K$\alpha$ emission in AGNs and the dependence of the spectral parameters on the X-ray luminosity and redshift.}
{We selected 2646 point sources from the 2XMM catalogue at high galactic latitude ($|BII|$ $>$ 25 degrees) and with the sum of EPIC-PN and EPIC-MOS 0.2-12 keV counts greater than 1000. Redshifts were obtained for 916 sources from the Nasa's Extragalactic Database. We excluded sources classified as HII regions, groups/clusters, star-forming/starburst galaxies. The final sample consists of 507 AGN. Individual source spectra have been summed in the observed frame to compute the integrated spectra in different redshift and luminosity bins over the range 0$<$z$<$5. Detailed analysis of these spectra has been performed using appropriately normalized background spectra and exposure time weighted response and ancillary files.} 
{We find that the narrow Fe K$\alpha$ line at 6.4 keV is significantly detected up to z=1. The line equivalent width decreases with increasing X-ray luminosity in the 2-10 keV band (``Iwasawa-Taniguchi effect''). The anticorrelation is characterized by the relation $\log(EW_{Fe}) = (1.66\pm0.09) + (-0.43\pm0.07) \log(L_{X,44})$, where $EW_{Fe}$ is the rest frame equivalent width of the neutral iron K$\alpha$ line in eV and $L_{X,44}$ is the 2-10 keV X-ray luminosity in units of $10^{44}$ erg s$^{-1}$. The equivalent width is nearly independent of redshift up to z$\sim$0.8 with an average value of 101$\pm$40 (rms dispersion) eV  in the luminosity range 43.5$\leq$logL$_X$$\leq$44.5. Our analysis also confirmed the hardening of the spectral indices at low luminosities implying a dependence of obscuration on luminosity.}
{We confirm that the neutral narrow Fe K$\alpha$ line is an almost ubiquitous feature of AGNs. We find compelling evidence for the ``Iwasawa-Taniguchi effect'' over a redshift interval larger than probed in any previous study. We detect no evolution of the average rest frame \eqw of \ka line with redshift.}


\maketitle
%

\section{Introduction}

Active galactic nuclei (AGNs) are believed to be powered by accretion onto supermassive black holes (SMBH) \citep{Salpeter1964, Lynden-Bell1969, Rees1984}. Most of the accretion power is released in the innermost region around the central SMBH \citep{Shakura1973, Pringle1981}, where the emerging spectrum is significantly affected by strong relativistic effects \citep{Page1974, Cunningham1975}. The Fe K$\alpha$ emission line is the most prominent spectral feature observed in the X-ray spectra of AGNs. The line profile serves as a powerful diagnostic tool to probe its origin and provides unique information on the dynamics of the emitting region \citep{Fabian1989, Laor1991}. 

The Fe K$\alpha$ emission was first identified as a common feature in the X-ray spectra of Seyfert galaxies observed by Ginga \citep{Pounds1990, Nandra1994}. Early studies of type 1 AGNs with ASCA revealed resolved, broad and asymmetric Fe K$\alpha$ lines in the majority of sources under study. The asymmetric lines were described as being originated from the accretion disc \citep{Nandra1997}. Recent Chandra and XMM AGNs observations show that narrow Fe K emission is a ubiquitous feature in their X-ray spectra  \citep{Yaqoob2004, Pounds2002}. However, unambiguous broad lines have been significantly detected and characterized only in a few Seyfert 1 galaxies, examples being MCG-6-30-15 \citep{Fabian2002}, Mrk 205 \citep{Reeves2001} and Mrk 509 \citep{Pounds2001}. The number of spectral counts in AGNs is a major constraint in characterizing the line properties, since an accurate modeling of the underlying continuum is essential to measure the line parameters. 

Several studies have been conducted to characterize the average Fe K$\alpha$ emission in local and distant AGNs. \citet{Nandra2007} performed a spectral analysis of a sample of 26 nearby Seyfert galaxies (z$<$0.05) observed by XMM-{\it Newton}. They found evidence for broad line emission in around 65\% of the sample with typical equivalent width (EW) being $\sim$77$\pm$16 eV, when fitted with a broad Gaussian. \citet{Gu2006} performed a similar analysis using a sample of 102 AGNs (including Seyfert galaxies, Radio Quiet Quasars) observed with XMM-{\it Newton} at z$<$0.5. They detected relativistic lines in about 25\% of the sample objects ($\simeq$50\% when the ``well exposed'' spectra with $\gtrsim$10000 net counts in the 2-10 keV band are considered). They also divided the sample in four equally populated luminosity classes and stacked residuals by modeling the continuum in the 2.5-15 keV energy range with an absorbed power law and Compton reflection component. They found that the average EWs derived from the stacked spectral residuals analysis are $\lesssim$150 eV for all luminosity classes. They reported that the strongest relativistic profiles may be present in low-luminosity (L$_X$$<$10$^{43}$ erg s$^{-1}$) AGNs.

\citet{Alina2005} derived an average rest frame spectrum of AGNs detected in a 770 ksec XMM-{\it Newton} observation of the Lockman Hole field. Using a sample of 104 X-ray sources (53 type 1 and 41 type 2 AGNs and 10 galaxies) covering 0$<$z$<$4.5 in redshift space, they reported a clear relativistic line in the average rest frame spectra with an EW of $\sim$560 eV and $\sim$460 eV for the type 1 and type 2 AGNs, respectively. \citet{Corral2008} computed the mean Fe emission from a large sample of more than 600 type 1 AGNs spanning a redshift range up to $\sim$3.5. They detected significant unresolved Fe K$\alpha$ emission line around 6.4 keV with an EW$\sim$90 eV and found no compelling evidence of any significant broad relativistic emission line in the average spectrum. 

\citet{Brusa2005} studied the average spectra of AGNs detected in the Chandra Deep Field North and South. They used a large sample of 352 spectroscopically identified AGNs and stacked X-ray counts in the observed frame in seven redshift bins over the range 0.5$<$z$<$4. The measured Fe K$\alpha$ line EWs are consistent with those reported by \citet{Alina2005}. However, a broad component was not required by the data. Based on these recent studies, the ubiquity of relativistically broadened Fe K$\alpha$ emission in AGNs and redshift dependence of the line strength are still unclear. 

Studying the relation between emission line strength (characterized by the EW) and continuum luminosity is very important because it unveils the possible association between the emitting regions of these two spectral components. \citet{Baldwin1977} first investigated the continuum luminosity dependence of the C IV $\lambda$1549 EW and discovered that the EW of the C IV line decreased with increasing UV luminosity, a phenomenon now known as the ``Baldwin Effect''. Since that time, confirming its existence and exploring its origin and evolution has become the subject of several studies \citep[e.g.][]{Dietrich2002, Green2001}. \citet{IT1993} reported the anticorrelation between the Fe K emission line EW and X-ray luminosity, based on Ginga observations of AGNs. This relation was designated as ``X-ray Baldwin effect'' or ``Iwasawa Taniguchi effect'' (hereafter IT effect). This trend was confirmed by later ASCA observations of broad iron lines \citep{Nandra1997}. In recent years, the existence of the IT effect for the narrow Fe K$\alpha$ emission lines in {\it type 1 AGNs} observed by XMM-{\it Newton} and Chandra has been well established \citep{Page2004, Zhou2005, Bianchi2007, Wu2009}. 

The most fundamental problem, however, is that the physical causes for the ``UV/optical Baldwin effect'' and ``IT effect'' still remain unknown. Several models have been proposed to account for these phenomena. \citet{Mush1984} suggested that the observed relation is due to a combination of decreasing ionization parameter and covering factor with increasing continuum luminosity. \citet{Korista1998} pointed out that the Baldwin effect is driven by a softening of the ionizing continuum and increasing metallicity towards higher luminosities. Other studies propose that the underlying physical parameters driving the Baldwin effect may be the Eddington ratio \citep{Baskin2004, Warner2004, Dong2009} and black hole mass \citep{Xu2008}. \citet{Nandra1997} attributed the IT effect to the presence of an ionized skin on the accretion disc, with the degree of ionization increasing with luminosity \citep{NayaB2000, NayaA2000}. 

 In the present study, we exploit the 2XMM catalogue to characterize the properties of the integrated spectrum to address the following key points : (i) ubiquity of the Fe K$\alpha$ emission line in AGNs, (ii) origin of the IT effect and (iii) evolution of the Fe K$\alpha$ \eqw with redshift. The conventional approach to improve the counting statistics in AGNs involves unfolding the instrumental response by appropriately modeling the continuum and then shifting and adding the individual spectra, which in turn, maximizes the S/N ratio around the iron line energy. However, the main drawback is a model dependent parameterization of the underlying continuum. We adopt an alternative approach, first implemented by \citet{Brusa2005}, in which source spectra are stacked in narrow redshift intervals.

The paper is organized as follows. In the following Section, the selection criteria and properties of the sample are discussed. Section 3 outlines the spectral analysis. The following Sections then present and discuss the results. Thoughout this work a cosmology with $\Omega_m$=0.27, $\Omega_\Lambda$=0.73 and H$_0$=70 km s$^{-1}$ Mpc$^{-1}$ is used.

\section{The Sample}
We use the sample derived from the 2XMM catalogue, the second comprehensive catalogue of serendipitous X-ray sources from the European Space Agency's (ESA) XMM-{\it Newton} observatory. The catalogue contains 246897 X-ray source detections which relate to 191870 unique X-ray sources \citep{XMMCAT}. We selected 2646 point sources\footnote {In the 2XMM catalogue point sources are characterized by the source extent parameter (EP$\_$EXTENT=0). It is worthwhile to point out that the source extent below 6$\arcsec$ is considered as a point source and the extent is re-set to zero.} satisfying the following selection criteria : (i) high galactic latitude ($|BII|$ $>$ 25 degrees) and (ii) sum of the EPIC-PN and EPIC-MOS counts in 0.2-12 keV  greater than 1000. We cross-correlated the list with the Nasa's Extragalactic Database and  obtained redshift information for 916 sources ($\sim$35\% of the sample).

 Sources in the 2XMM catalogue are assigned `automated' and `manual' quality warning flags. These flags provide essential information about the possible problematic issues such as proximity to a bright source, a location within an extended source emission, insufficient detector coverage of the point spread function of the detection etc. Additionally, the catalogue gives a summary flag which combines the automated and manual quality warning flags. The summary flag, with values in the range 0-4, indicates the overall quality of each detection ordered by increasing severity. For a detailed description of possible values of the summary flag see \citet{XMMCAT}.

The summary flags of the 916 sources were carefully checked. To obtain the cleanest possible sample, we retained  the sources with `0' summary flags and removed the remaining sources from the sample. A source is assigned `0' summary flag when none of the automated or manual quality flags is set to True. We  also used optical spectroscopic information for further screening of the sample and excluded all sources classified as HII regions, in groups/clusters and starburst/star forming galaxies. The final sample consists of 507 point sources.

\begin{figure}
\begin{center}
\centering
\resizebox{70mm}{!}{\includegraphics{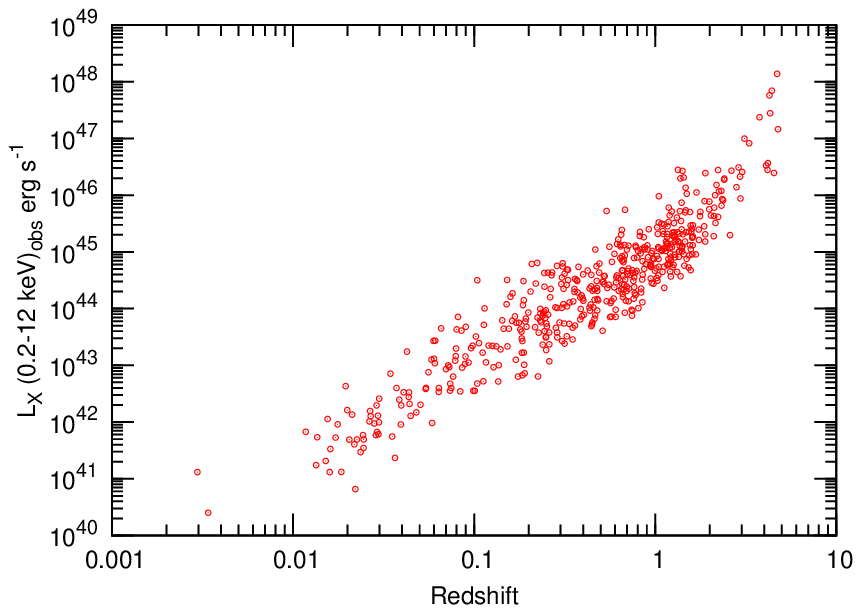}}
\resizebox{70mm}{!}{\includegraphics{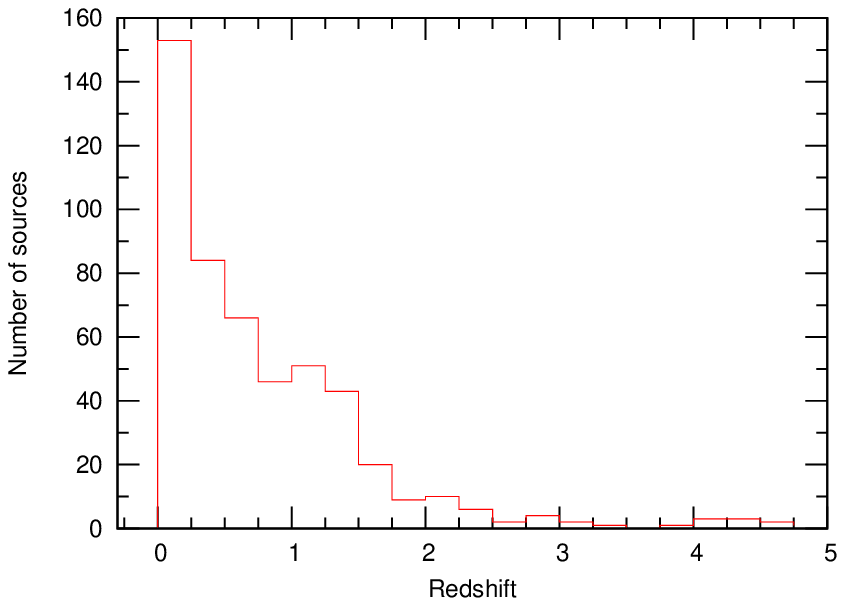}}
\caption{Luminosity-redshift distribution (top) and redshift distribution (bottom) of the sources in the final sample.}
\label{fig: L-zdist}
\end{center}
\end{figure}

We then grouped the sources according to their redshifts in 13 redshift bins in the range 0$<$z$<$5. The choice of bin sizes is driven by a trade-off between the sum of net counts in each bin and to avoid too wide redshift bin preventing the detection or smearing the narrow line feature. The 0.2-12 keV flux (observed frame) given in the 2XMM catalogue  was used to calculate the luminosity of the sources. In Table \ref{table: source-stat}, we report for each redshift bin: the total number of sources, EPIC-PN counts in the 2-10 keV rest frame and the total number of luminosity bins in which the sources were further splitted. The distribution in luminosity and redshift of the final sample is shown in Figure \ref{fig: L-zdist}. 

\begin{table}
\centering
\caption{Statistics of the sources}
\begin{tabular}{*{4}{lll}}
\hline\hline
 \T Redshift & Total & PN net counts & Total \\
 bin & sources & \B (2-10 keV) & Lbins \\ \hline
 \T 0.0-0.2 & 130 & 78857 & 6\\
 0.2-0.3 & 42 & 33896 & 2\\
 0.3-0.4 & 37 & 15594  & 1\\
 0.4-0.6 & 47 & 24349  & 2\\
 0.6-0.8 & 60 & 35550  & 2\\
 0.8-1.0 & 33 & 14502  & 2\\
 1.0-1.3 & 60 & 30263 & 2 \\
 1.3-1.7 & 52 & 35735 & 2\\
 1.7-2.1 & 12 & 5401 & 1\\
 2.1-2.5 & 15 & 9352 & 1\\
 2.5-3.0 & 6 & 5098 & 1\\
 3.0-4.0 & 4 & 6717 & 1\\
 4.0-5.0 & 9 & 12611 & 1\\ \hline
\end{tabular}
\label{table: source-stat}
\end{table}

\section{Data Analysis}
For each object, source specific products such as the source and background spectra, images in the different energy bands and response files were retrieved from the XMM-{\it Newton} Science Archive (hereafter XSA). Keeping in mind the limitations generally associated with the automatic detection process, several safety checks were carried out before using the archival products. 

First, we examined the quality of the archival products by extracting source, background spectra, and response matrices from the observation data files (ODF) for a few sources using the XMM-{\it Newton} Science Analysis Software (SAS, ver. 7.1.0). The sources were arbitrarily selected with '0' quality summary flag and varying EPIC-PN counts spanning a range of $\sim$300 to 5000 in the 0.2-12 keV band. We cross-compared these ODF extracted spectra with the pipeline processed archival spectra retrieved from the XSA. We found a very good agreement between these spectra for all sources under study. 

Second, we visually inspected the EPIC-PN images of the sources in the final sample to check if there are additional sources in the background annulus. In a few images, background annuli were found to have additional sources. We excluded all the archival products of these crowded field images from the stacking analysis. These exercises ensured us the goodness and reliable quality of archival products. Only EPIC-PN data has been used in our work. 

\subsection{Spectral fitting}
Individual source spectra have been stacked in the observed frame in 13 redshift bins over the range 0$<$z$<$5. To avoid the contamination of the stacked spectrum by sources covering a too wide range of luminosities in a redshift bin, sources are further splitted into luminosity intervals. The choice of luminosity intervals associated to a redshift bin is driven by the number of sources in that redshift bin. For this reason, sources in all redshift bins at z$>$1.7 are kept in only one luminosity interval. In all luminosity interval(s) of a redshift bin, the contribution of the single source counts to the total counts (0.2-12 keV) was checked before stacking spectra. The sources with more than 40$\%$ contribution were removed from the luminosity interval. Appropriate background, response and ancillary files were created using the FTOOLS routines mathpha, addrmf and addarf, respectively. The response and ancillary files were weighted for the exposure time. We refer to these count-rate averaged X-ray spectra as the ``real stacked spectra''.

\begin{figure*}
\begin{center}
\begin{tabular}{cc}
\resizebox{70mm}{!}{\includegraphics{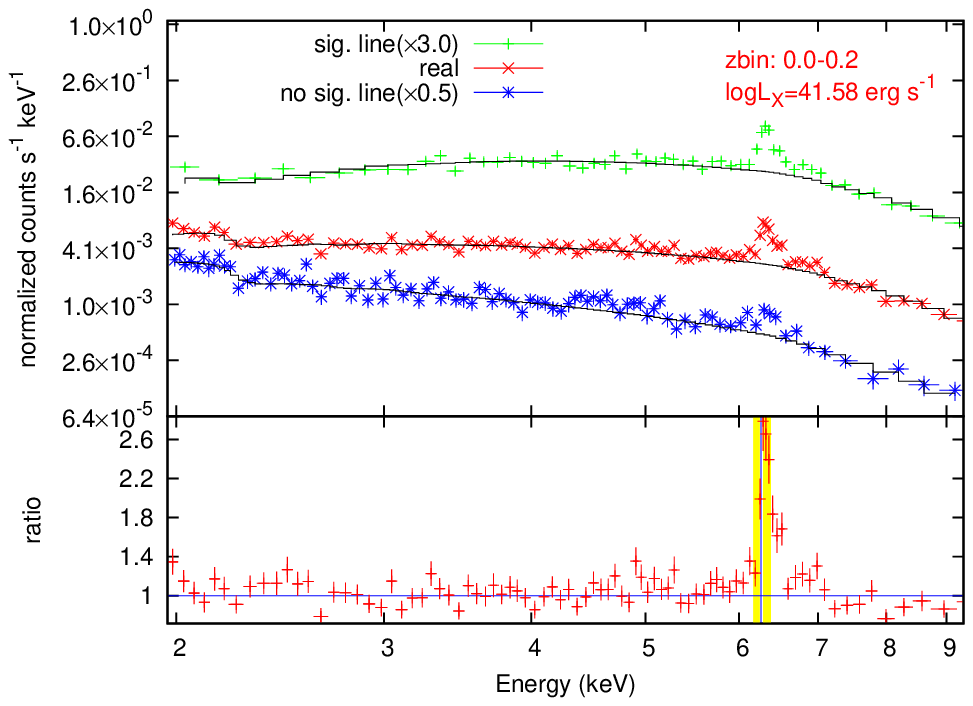}} &
\resizebox{70mm}{!}{\includegraphics{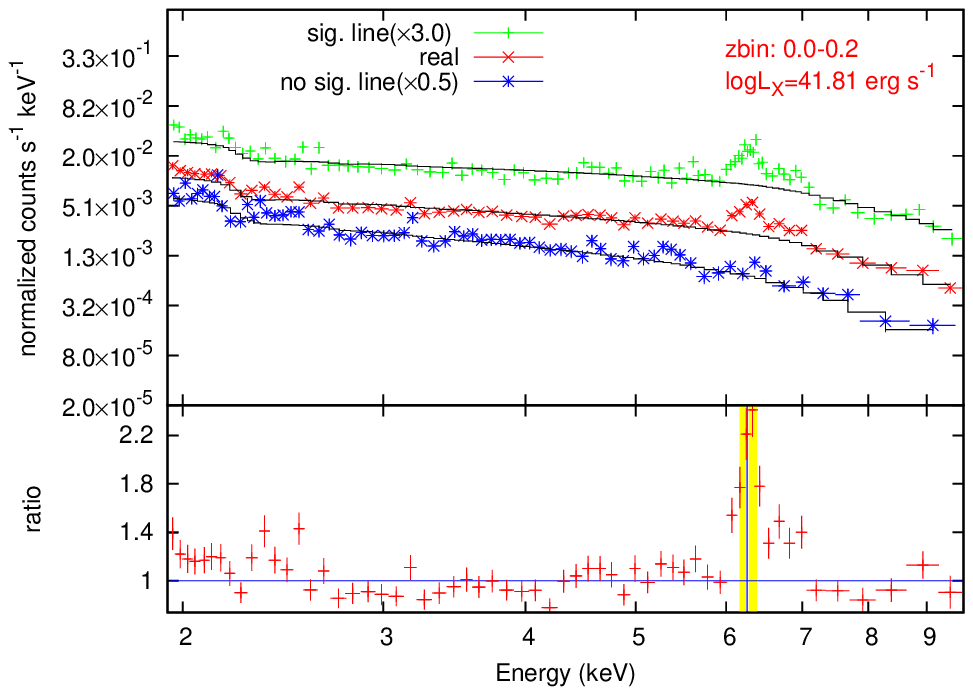}} \\
\resizebox{70mm}{!}{\includegraphics{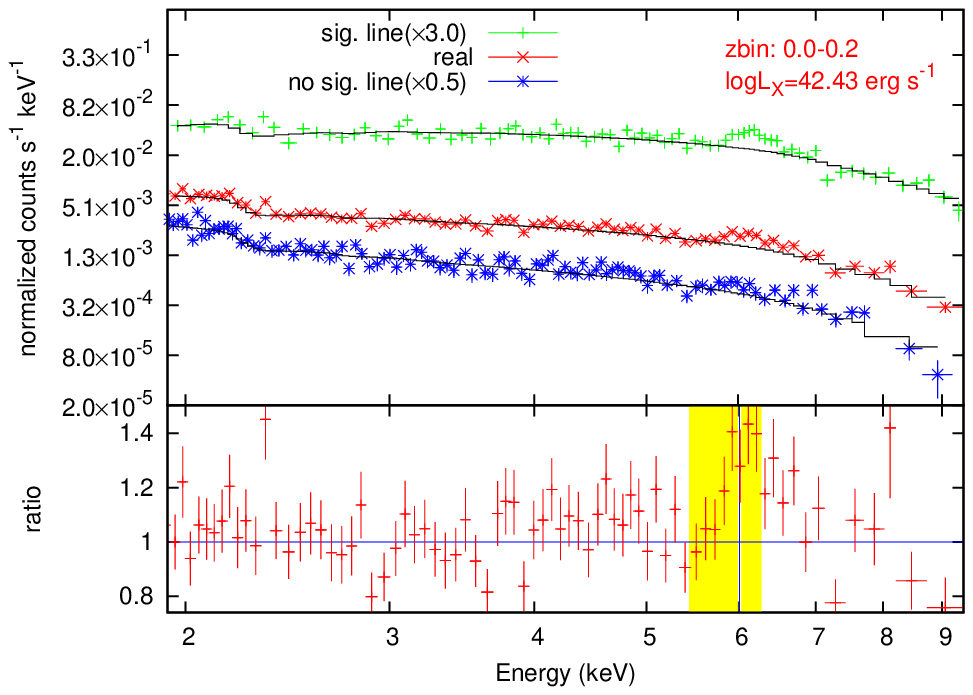}} &
\resizebox{70mm}{!}{\includegraphics{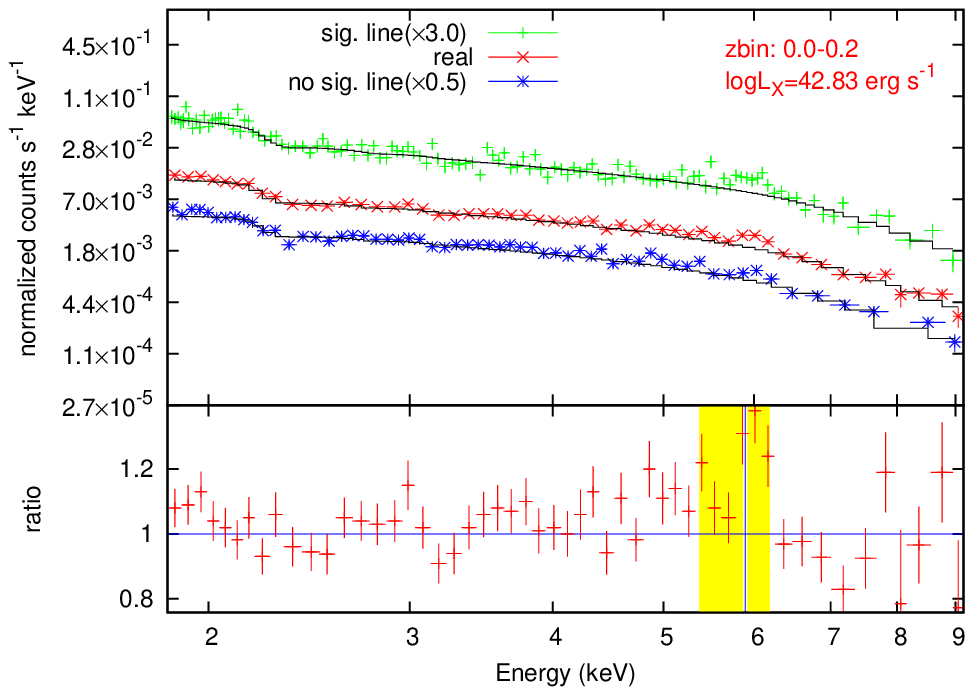}} \\
\resizebox{70mm}{!}{\includegraphics{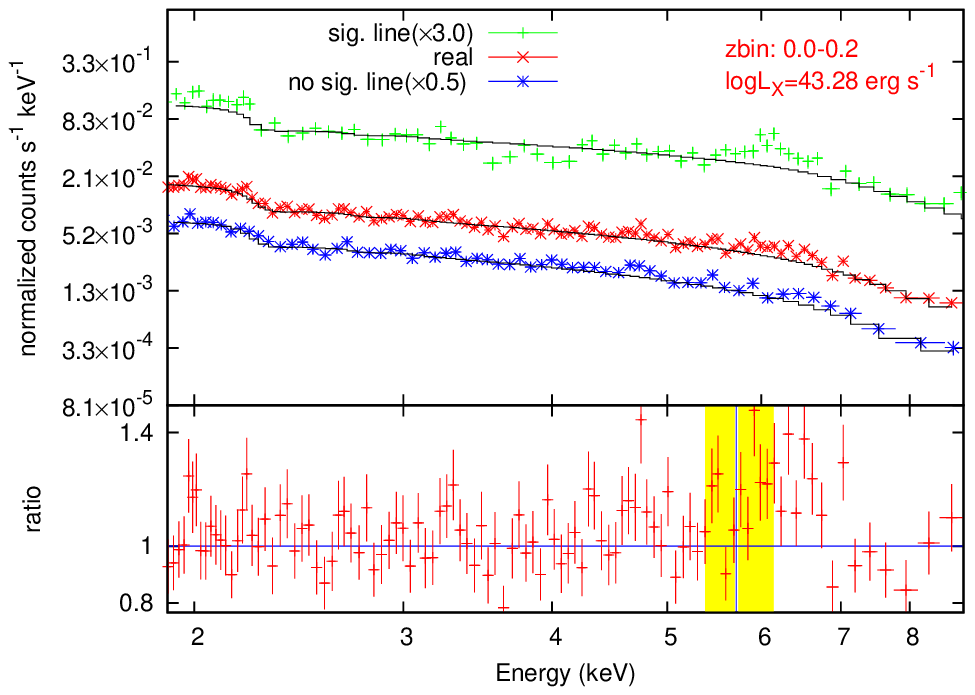}} &
\resizebox{70mm}{!}{\includegraphics{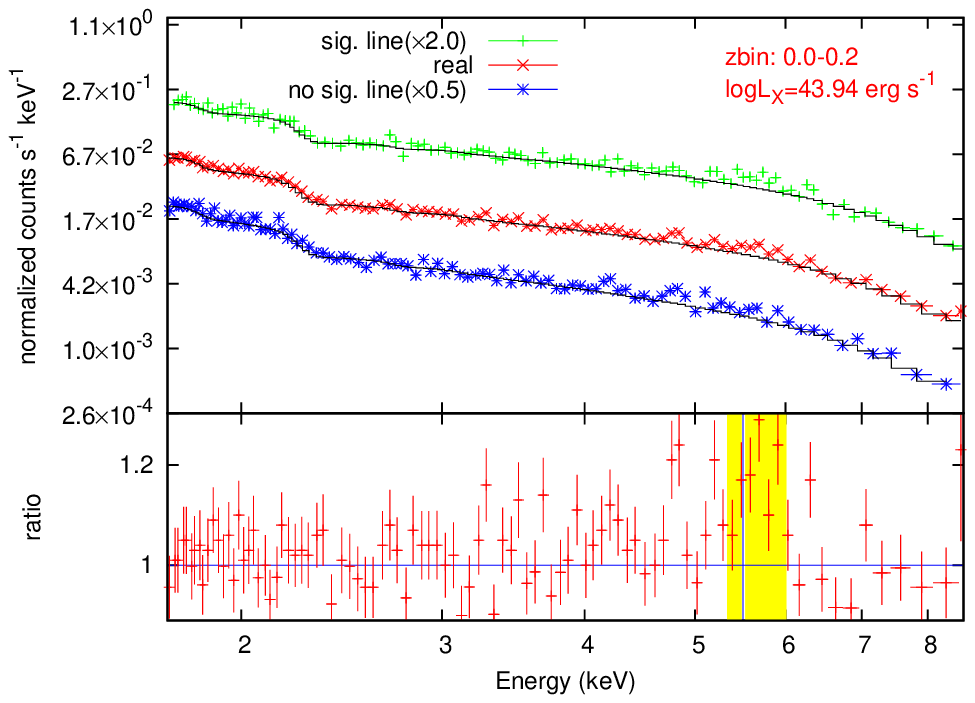}} \\
\resizebox{70mm}{!}{\includegraphics{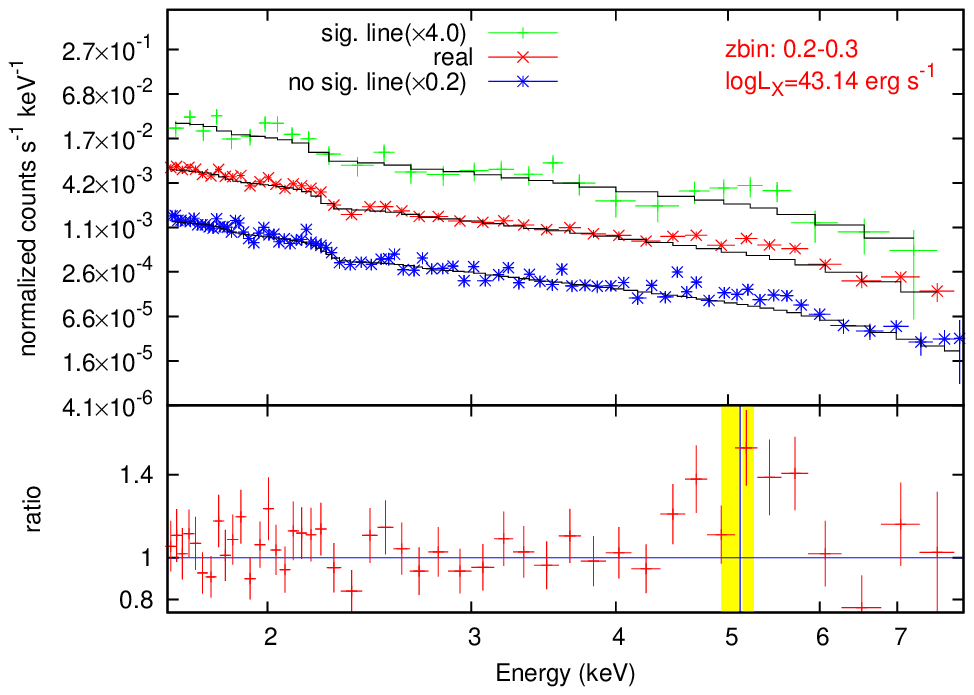}} &
\resizebox{70mm}{!}{\includegraphics{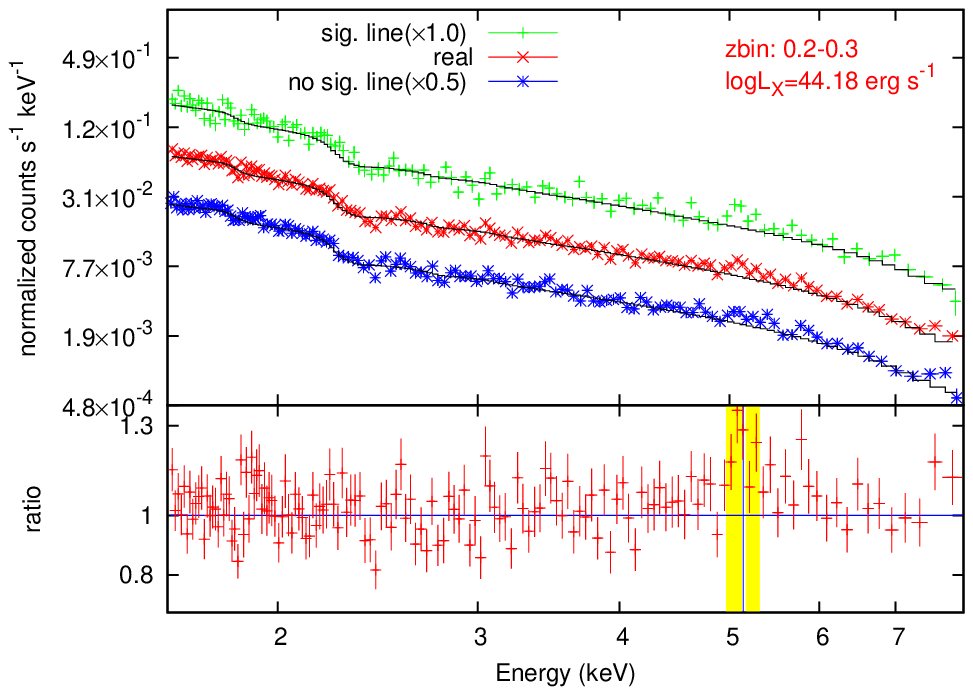}} \\
\end{tabular}
\caption{Top panels in each figure show spectral fits (mo wabs(pow)) to three stacked spectra in the quoted redshift bin, from top to bottom, sources with significant Fe K$\alpha$ line detection, real stacked spectra and sources with undetected Fe K$\alpha$ line. For visual clarity, the stacked spectra of sources with and without significant Fe K$\alpha$ line detection have been rescaled by the factor listed in parenthesis. In the bottom panels Data/Model ratios are shown for the real stacked spectra. The vertical line is drawn at the expected position for the redshifted 6.4 keV Fe k$\alpha$ line, while the shaded region encompasses the bin width reported in Table 2. The average rest frame luminosities (2-10 keV) obtained for the best fit model (a single absorbed power law and a narrow redshifted Gaussian line) are also quoted for the real stacked spectra.}
\label{fig: ss-set1}
\end{center}
\end{figure*}

\begin{figure*}
\begin{center}
\begin{tabular}{cc}

\resizebox{70mm}{!}{\includegraphics{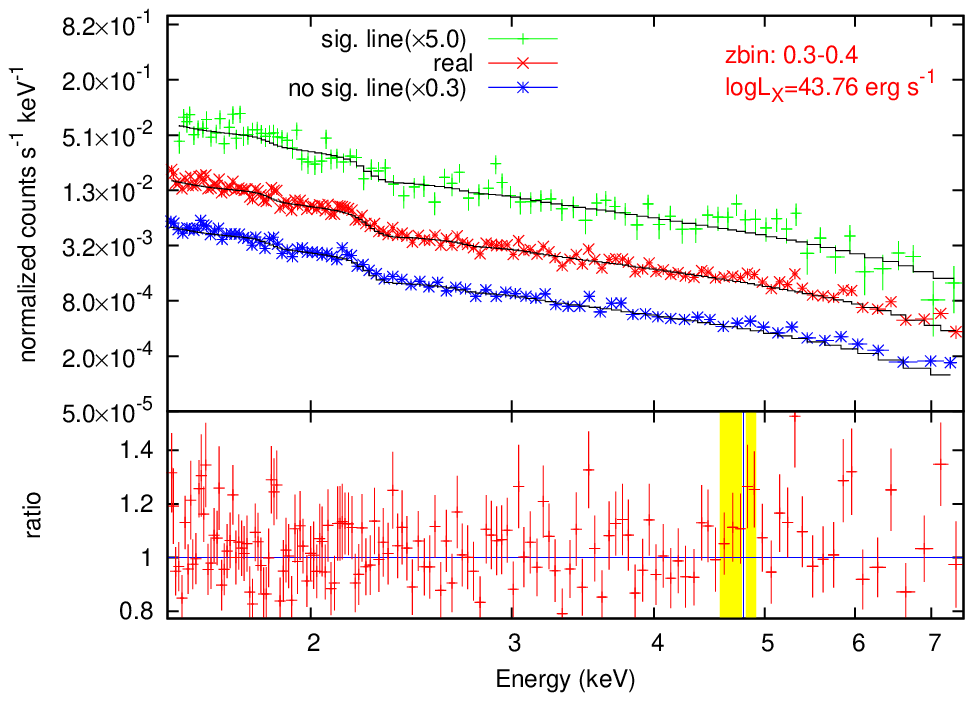}} &
\resizebox{70mm}{!}{\includegraphics{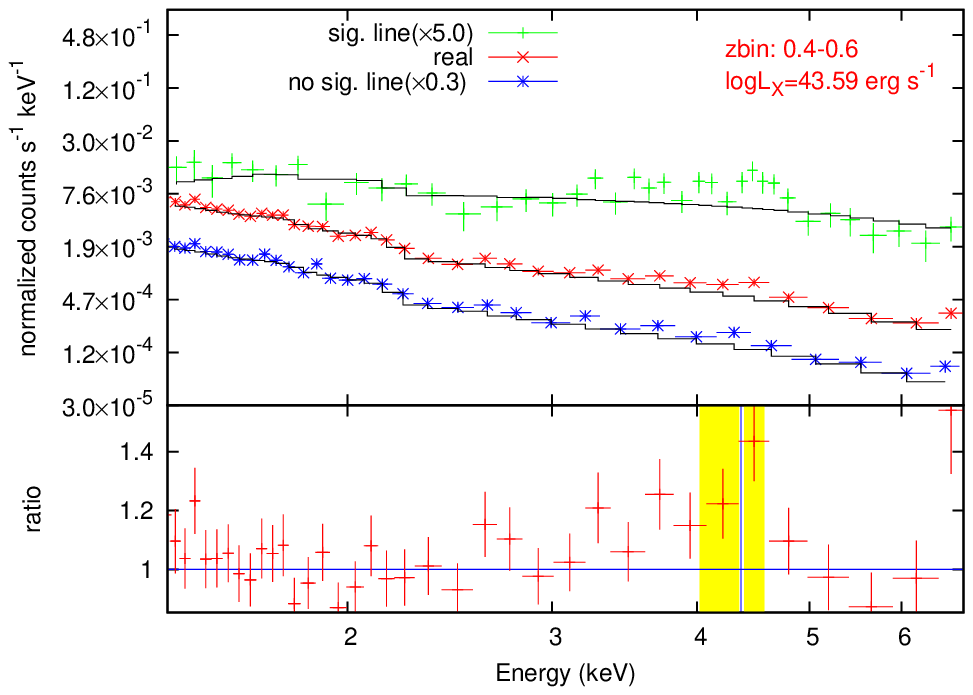}} \\
\resizebox{70mm}{!}{\includegraphics{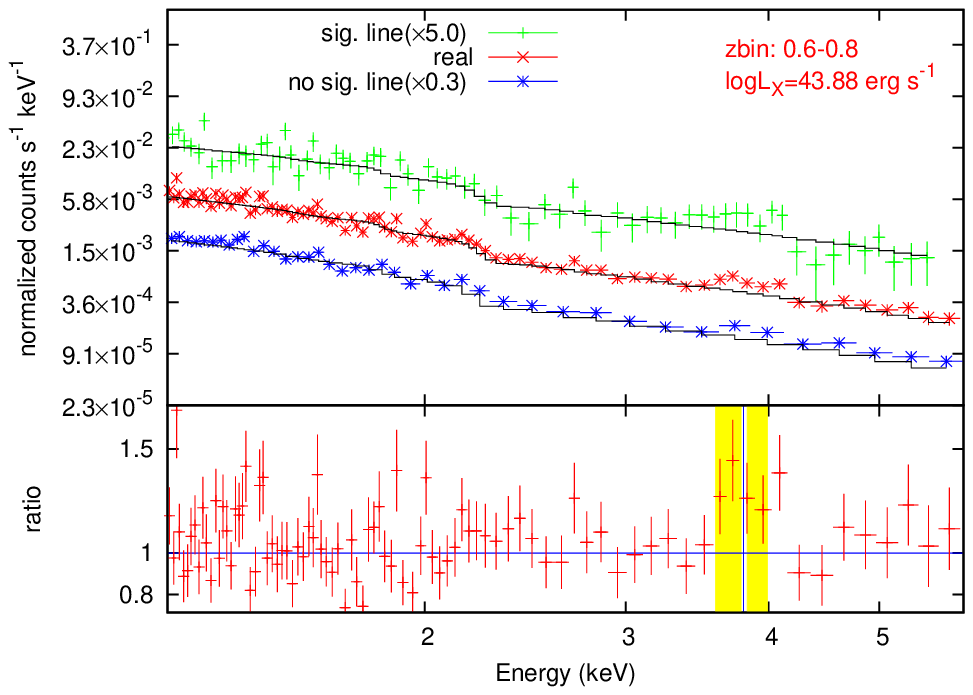}} &
\resizebox{70mm}{!}{\includegraphics{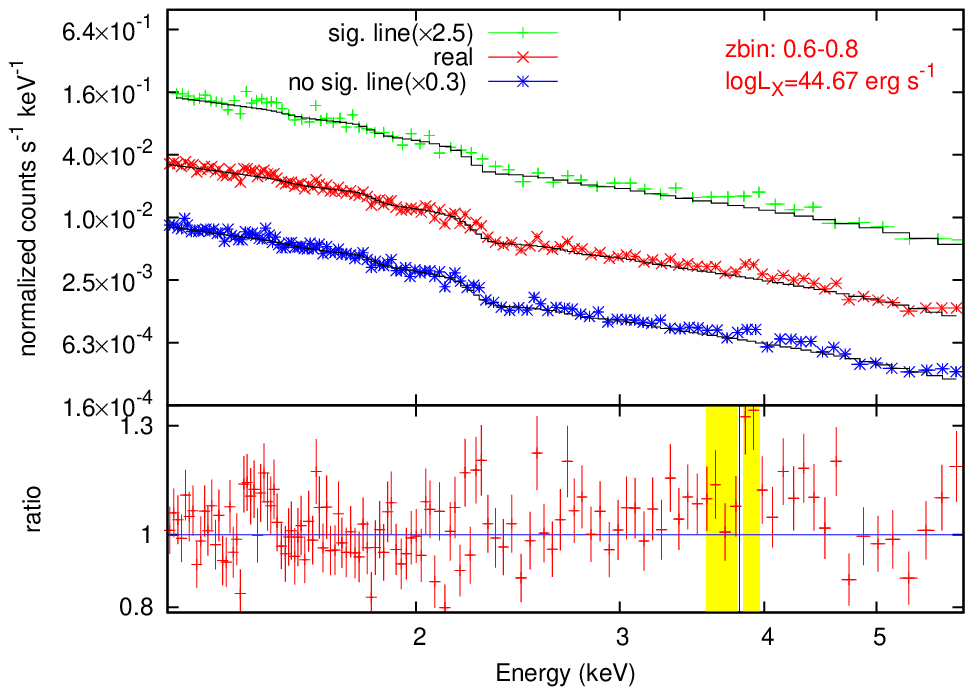}} \\
\end{tabular}
\caption{Representation of the data in this figure is the same as in Figure \ref{fig: ss-set1}.}
\label{fig: ss-set2}
\end{center}
\end{figure*}

\begin{table*}
\caption{Results of spectral fitting in different redshift bins.}
\begin{tabular}{*{11}{c}} \hline\hline
\T z$_{median}$ & E(K$\alpha$) & Binwidth & Total & Spectral & log L$_X$(2-10 keV) & $\Gamma$ & Eqw & $\chi^2$/dof  & $\chi^2$/dof & F-prob \\
& (keV)  & (keV) & sources & counts & erg s$^{-1}$ &  & (eV) & wabs(pow+zgauss) & wabs(pow) & \\ 
 (1) & (2) &(3) &(4)& (5)& (6)& (7)& (8) & (9) & (10) & \B (11)\\\hline
\multicolumn{11}{c}{\T zbin: 0.0-0.2 \B}\\ \hline
\T 0.02 & 6.26 & 0.21 & 20(8) & 10440 & 41.58$^{+0.05}_{-0.11}$ & 0.93$^{+0.08}_{-0.07}$ & 398$^{+54}_{-53}$ & 491.9/435 & 642.5/437 & \B $>$99.999\\
0.03 & 6.22 & 0.28 & 20(9) & 8421 & 41.81$^{+0.04}_{-0.14}$ & 1.04$^{+0.06}_{-0.05}$ & 455$^{+65}_{-66}$ & 420.8/351 & 544.9/353 & \B $>$99.999\\
0.06 & 6.01 & 0.83 & 23(4) & 9433 & 42.43$^{+0.05}_{-0.12}$ & 1.06$^{+0.09}_{-0.08}$ & 163$^{+54}_{-52}$ & 373.4/385 & 399.3/387 & \B $>$99.999\\
0.09 & 5.89 & 0.81 & 21(6) & 13986 & 42.83$^{+0.05}_{-0.10}$ & 1.52$^{+0.10}_{-0.09}$ & 105$^{+66}_{-31}$ & 508.9/518 & 527.7/520 &\B 99.991\\
0.12 & 5.71 & 0.75 & 22(3) & 11932 & 43.28$^{+0.05}_{-0.09}$ & 1.20$^{+0.08}_{-0.08}$ & 105$^{+45}_{-46}$ & 474.1/472 & 488.3/474 & \B 99.906\\ 
0.16 & 5.51 & 0.67 & 24(6) & 24645 & 43.94$^{+0.03}_{-0.06}$ & 1.37$^{+0.05}_{-0.04}$ & 75$^{+30}_{-29}$ & 712.9/737 & 729.3/739 & \B 99.976\\ \hline 
\multicolumn{11}{c}{\T zbin: 0.2-0.3 \B}\\ \hline
\T 0.25 & 5.12 & 0.32 & 21(2) & 6415 & 43.14$^{+0.05}_{-0.31}$ & 1.88$^{+0.12}_{-0.07}$ & 201$^{+110}_{-109}$ & 259.5/260 & 268.3/262 & \B 98.717\\
0.24 & 5.14 & 0.34 & 21(2) & 27481 & 44.18$^{+0.03}_{-0.06}$ & 1.72$^{+0.03}_{-0.02}$ & 79$^{+31}_{-30}$ & 728.8/708 & 746.2/710 & \B 99.976\\ \hline
\multicolumn{11}{c}{\T zbin: 0.3-0.4 \B}\\ \hline
\T 0.34 & 4.79 & 0.34 & 37(5) & 15594 & 43.76$^{+0.03}_{-0.09}$ & 1.82$^{+0.05}_{-0.03}$ & 69$^{+49}_{-49}$ & 439.6/477 & 444.8/479 & \B 94.026\\ \hline
\multicolumn{11}{c}{\T zbin: 0.4-0.6 \B}\\ \hline
\T 0.47 & 4.37 & 0.54 & 23(2) & 7099 & 43.59$^{+0.04}_{-0.17}$ & 1.71$^{+0.08}_{-0.06}$ & 144$^{+95}_{-93}$ & 281.5/274 & 287.7/276 & \B 94.933\\
0.51 & 4.24 & 0.48 & 24 & 17250 & 44.41$^{+0.04}_{-0.06}$ & 1.74$^{+0.04}_{-0.06}$ & $<$98 & 475.3/485 & 479.4/487 & \B 87.255\\ \hline
\multicolumn{11}{c}{\T zbin: 0.6-0.8\B}\\ \hline
\T 0.68 & 3.80 & 0.40 & 30(4) & 9296 & 43.88$^{+0.04}_{-0.13}$ & 1.90$^{+0.07}_{-0.07}$ & 163$^{+87}_{-96}$ & 297.8/330 & 305.7/332 & \B 98.672\\
0.68 & 3.81 & 0.40 & 30(3) & 26254 & 44.67$^{+0.02}_{-0.05}$ & 1.83$^{+0.03}_{-0.03}$ & 93$^{+39}_{-39}$ & 602.3/561 & 616.9/563 & \B 99.881\\ \hline
\multicolumn{11}{c}{\T zbin: 0.8-1.0 \B}\\ \hline
\T 0.89 & 3.39 & 0.33 & 17 & 7142 & 44.17$^{+0.03}_{-0.12}$ & 1.90$^{+0.09}_{-0.07}$ & $<$223 & 254.1/262 & 258.4/264 & \B 88.928\\
0.91 & 3.36 & 0.32 & 16 & 7360 & 44.69$^{+0.03}_{-0.10}$ & 1.88$^{+0.06}_{-0.06}$ & $<$135 & 263.8/264 & 265.1/266 & \B 47.559\\ \hline
\multicolumn{11}{c}{\T zbin: 1.0-1.3 \B}\\ \hline
\T 1.16 & 2.97 & 0.38 & 30 & 12833 & 44.45$^{+0.02}_{-0.08}$ & 2.06$^{+0.08}_{-0.05}$ & $<$183 & 325.9/354 & 329.8/356 &\B 88.076\\
1.18 & 2.94 & 0.36 & 30 & 17430 & 45.02$^{+0.02}_{-0.05}$ & 1.85$^{+0.05}_{-0.04}$ & $<$51 & 423.0/417 & 423.0/418 & \B 5.595\\ \hline
\multicolumn{11}{c}{\T zbin: 1.3-1.7 \B}\\ \hline
\T 1.44 & 2.63 & 0.38 & 27 & 12255 & 44.51$^{+0.04}_{-0.07}$ & 1.99$^{+0.13}_{-0.10}$ & $<$174  & 358.9/352 & 361.5/354 &\B 71.753\\
1.46 & 2.60 & 0.38 & 25 & 23180 & 45.31$^{+0.02}_{-0.04}$ & 1.89$^{+0.07}_{-0.04}$ & $<$110 & 468.6/441 & 473.4/443 & \B 89.606\\ \hline
\multicolumn{11}{c}{\T zbin: 1.7-2.1 \B}\\ \hline
\T 1.89 & 2.22 & 0.23 & 12 & 5401 & 45.03$^{+0.04}_{-0.10}$ & 1.69$^{+0.17}_{-0.08}$ & $<$193 & 199.1/200 & 201.9/201 & \B 90.608\\ \hline
\multicolumn{11}{c}{\T zbin: 2.1-2.5 \B}\\ \hline
\T 2.22 & 1.99 & 0.18 & 15 & 9352 & 45.38$^{+0.04}_{-0.05}$ & 1.92$^{+0.14}_{-0.12}$ & $<$48 & 291.2/279 & 291.2/280 & \B 0.000\\ \hline
\multicolumn{11}{c}{\T zbin: 2.5-3.0 \B}\\ \hline
\T 2.88 & 1.65 & 0.16 & 6 & 5098 & 45.56$^{+0.04}_{-0.06}$ & 2.00$^{+0.20}_{-0.13}$ & $<$115  & 177.5/190 & 178.0/191 & \B 50.559\\\hline
\multicolumn{11}{c}{\T zbin: 3.0-4.0 \B}\\ \hline
\T 3.31 & 1.49 & 0.25 & 4 & 6717 & 46.18$^{+0.04}_{-0.05}$ & 1.66$^{+0.16}_{-0.14}$ & $<$106  & 224.9/222 & 225.6/223  & \B 59.089\\ \hline
\multicolumn{11}{c}{\T zbin: 4.0-5.0 \B}\\ \hline
\T 4.31 & 1.21 & 0.14 & 9 & 12611 & 46.32$^{+0.03}_{-0.03}$ & 1.76$^{+0.13}_{-0.11}$ & $<$83  & 275.2/284 & 276.5/285 & \B 74.598\\ \hline
\end{tabular}
\label{table:fit-results}
 Notes: 
 Col. (1) : Median redshift of the bin.
 Col. (2) : Expected energy of the neutral Fe K$\alpha$ line.
 Col. (3) : Observed frame bin width at the 6.4 keV position, defined as $\triangle$E= 6.4/1+z$_{min}$-6.4/1+z$_{max}$. Comparatively large bin widths in the redshift bin 0.0-0.2 are due to the variation in z$_{min}$ and z$_{max}$ introduced by splitting the sources in different luminosity bins. 
 Col. (4) : Total number of stacked sources in different luminosity bins in each redshift bin (in parenthesis number of sources with significant Fe K$\alpha$ line).
 Col. (5) : Total net counts in the rest frame 2-10 keV band.
 Col. (6) : Rest frame X-ray luminosity measured in the 2-10 keV band.
 Col. (7) : Power law photon index quoted over the 2-10 keV band; errors are reported at the 90\% confidence level for one interesting parameter.
 Col. (8) : Rest frame equivalent width of the Fe K$\alpha$ line.
 Col. (9) : $\chi^2$ and degrees of freedom using a single absorbed powerlaw and a redshifted narrow Gaussian line.
 Col. (10) : $\chi^2$ and degrees of freedom using a single absorbed powerlaw.
 Col. (11) : The F-test probability of the line detection, for F-prob$\leq$90\% the upper limits on the EW are quoted at the 90\% confidence level. 
\end{table*}

The real stacked spectra were rebinned to have at least 20 counts per bin and analyzed using XSPEC \citep[ver. 12.4.0;][]{Arnaud1996}. In all redshift and their associated luminosity bins, the real stacked spectra were initially fitted with a single absorbed power law in the rest frame 2-10 keV band. In about half of the sample we found a significant excess above the power law continuum in the energy range from 6.4/(1+z$_{max}$) to 6.4/(1+z$_{min}$) keV, where z$_{min}$ and z$_{max}$ are the bin boundaries. To investigate the presence of a Fe emission line, we added a narrow redshifted Gaussian component (model wabs(pow+zgauss)) with the line width fixed at $\sigma$=0.1 keV and the redshift fixed at the median of the bin. We noticed that the fit improved in all luminosity bins, but with varying significance (see Table \ref{table:fit-results}). An F-test has been performed to test the significance of the narrow redshifted Gaussian line component. Given the well known limitations of the F-test in determining the precise significance of a spectral line \citep[see][]{Protassov2002}, we ran simulations in order to quantify the Fe line significance in the real stacked spectra (see Subsection \ref{simulations}).

In all redshift and luminosity bins at z$<$1.7, the peak energy of the Fe line is allowed to vary, whereas, at z$>$1.7, the centroid of the Fe line was fixed at 6.4 keV to constrain the fit. Whenever the addition of the line component was not significant (its significance $<$90\% according to an F-test), an estimate of the upper limit to the rest frame EW of the neutral Fe emission line was inferred. Errors on the Fe line EW were calculated based on the error in the line normalization considering the 90\% confidence interval. The fitting results are summarized in Table \ref{table:fit-results}.

The first part of Table \ref{table:fit-results} lists for each redshift bin and its associated luminosity bins, the expected positions of the Fe K$\alpha$ line, the bin width, the source statistics (number of sources and number of counts) and the average 2-10 keV luminosity. In the second part of Table \ref{table:fit-results}, results of the spectral fitting in each redshift bin (splitted into different luminosity bins) are presented, namely the photon index of the underlying power-law continuum ($\Gamma$), the EW of the Gaussian line component, $\chi^2$ and degrees of freedom associated with an absorbed powerlaw and with an additional redshifted narrow line component. The F-test probability of the line detection is also quoted.

\subsection{Contribution from single sources}
 In redshift and luminosity bins where a significant Fe K$\alpha$ line is detected in the real stacked spectrum, we performed spectral analysis of each source spectrum to check for the presence of a significant Fe K$\alpha$ line. We first grouped each source spectrum to have minimum 20 counts per bin. As our main interest is to identify spectra with significant Fe K$\alpha$ line, we restricted our spectral analysis in the 2-10 keV rest frame energy range. For simplicity we adopt the same continuum model for all spectra as for the real stacked spectrum. We then added a narrow ($\sigma$=0.1 keV) redshifted Gaussian line with the redshift fixed for each source. We performed an F-test to quantify the significance of the Fe K$\alpha$ line in each spectrum. We considered significant Fe K$\alpha$ detection in the spectrum if the F-prob of the line detection was $\geq$90\%. The number of sources found to have significant Fe K$\alpha$ line are quoted in parenthesis in Table \ref{table:fit-results}.

 We then investigated the Fe K$\alpha$ contribution from spectra with and without significant Fe K$\alpha$ lines by creating stacked spectra of both subsamples. These stacked spectra were analyzed with the same models as that used for the real stacked spectra and fit parameters were derived. The stacked spectra of sources without significant Fe K$\alpha$ line detection show the presence of Fe K$\alpha$ line with varying significance depending on the redshift and luminosity bin. Typical Fe K$\alpha$ EWs of these spectra span a range from  $\sim$200 eV in the lowest redshift bin to $\sim$100 eV in the redshift bin 0.6-0.8. In the top panels of Figures \ref{fig: ss-set1} and \ref{fig: ss-set2}, we show spectral fits with respect to a single absorbed power law to three stacked spectra, from top to bottom, sources with significant Fe K$\alpha$ line detection, real stacked spectra and sources with undetected Fe K$\alpha$ line. In the bottom panels Data/Model ratios are shown for the real stacked spectra.

\subsection{Simulations} \label{simulations}
We performed extensive simulations to assess the accuracy of the stacking procedure. We created ``simulated stacked spectra'' in redshift and luminosity bins where a significant Fe K$\alpha$ line is detected in the real stacked spectrum. To create the simulated stacked spectrum we considered separately the contribution from sources with and without significant Fe K$\alpha$ line. For the sources with significant Fe K$\alpha$ line, the best fit parameters retrieved from the individual spectral analysis were given as input parameters. For the sources without significant Fe K$\alpha$ line, the average values measured from their stacked spectrum (blue spectrum in Figures \ref{fig: ss-set1} and \ref{fig: ss-set2}) were used as input parameters. We produced 1000 simulated stacked spectra in each of the redshift and luminosity bins. Identical response and effective area files were used for the real and simulated stacked spectra.

 Each simulated stacked spectrum was subsequently fitted with a single absorbed powerlaw and a redshifted Gaussian line with the line width and redshift fixed at 0.1 keV and z$_{median}$, respectively. Then, histograms of all the spectral fit parameters were created and fitted with a Gaussian function to derive the mean and standard deviations of their distributions. Comparison of the spectral parameters of the real (see Table \ref{table:fit-results}) and simulated stacked spectra yielded a good agreement between them, in all cases within the statistical errors. Thus, we quantified that the EW of the Fe K$\alpha$ derived from the spectral analysis of the real stacked spectrum represents the average of the EWs of the Fe K$\alpha$ lines of the individual spectra comprising the real stacked spectrum.

\section{Results and Discussion}

Here we discuss the results of our work in different redshift and luminosity bins. At low redshifts (z$\leq$0.4) F-test probability of the Fe K$\alpha$ line detection is greater than 90\% in all luminosity bins. In the redshift range 0.4$<$z$\leq$0.8 the average F-test probability of the Fe K$\alpha$ line detection is $\sim$95\% in the lowest luminosity bin. An upper limit on the Fe K$\alpha$ EW was derived for the non-significant line detections in all redshift and luminosity bins at z$>$0.8. 

One key advantage of our sample with respect to previous published works is that it includes higher statistics covering a wide range of luminosities at various redshifts. We can therefore address the separate redshift and luminosity dependences of the Fe line strength.

\begin{figure}
\centering
\resizebox{80mm}{!}{\includegraphics{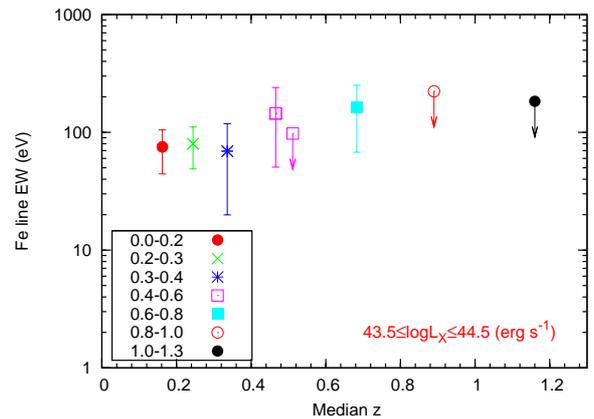}}
\caption{Rest frame equivalent width as a function of redshift.}
\label{fig: non-evolution}
\end{figure}

\subsection{Evolution of the Fe line equivalent width with redshift} {\label{subsec-line-z-dependence}}
First, we searched for a dependence of the Fe K$\alpha$ line intensity on cosmic time (redshift). We selected the narrow luminosity range of 43.5$\leq$logL$_X$$\leq$44.5 to cover a redshift range as large as possible for an almost constant luminosity. This sample contains 206 sources grouped in 7 redshift bins in the range 0$<$z$<$1.3 and an average luminosity of logL$_X$(2-10 keV)=44.04$\pm$0.30 (rms dispersion). We find that the intensity of the Fe K$\alpha$ line does not change significantly up to z$\sim$0.8 as displayed in Figure \ref{fig: non-evolution}, where the measured EWs with their associated errors and upper limits are plotted. The upper limits at z$>$0.8 are also consistent with a non evolution. The non evolution of the Fe K$\alpha$ line intensity can be interpreted as constant Fe abundance with redshift \citep{Brusa2005}. Studies of quasar elemental abundances (derived from broad optical emission lines) also reported a similar (non evolutionary) trend of metallicity with redshift up to z$\sim$4.5 \citep{Hamann2007}. In contrast, \citet{Balestra2007} found significant evidence of a decrease in the average Fe abundance of the intra-cluster medium as a function of redshift, at least up to z$\simeq$0.5. The comparison of these two results implies a different evolution of the Fe abundance with redshift in these two very different systems (AGNs and ICM) and/or very different scales (pc versus Mpc).

\subsection{The IT effect} 

\begin{figure*}
\begin{center}
\includegraphics[width=16cm, angle=0]{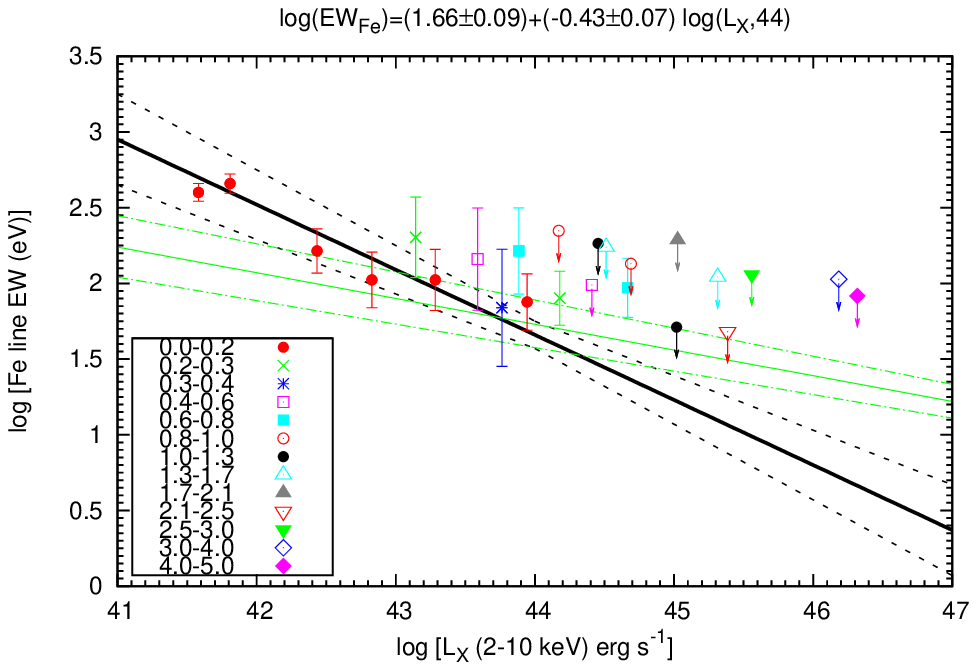}
\end{center}
\caption{Rest frame equivalent width of the narrow Fe K$\alpha$ line ($\sigma$=0.1 keV) in different redshift intervals plotted against the X-ray luminosity in the 2-10 keV band. The solid black line indicates the best fit to all the detections including upper limits. The fit relation is reported on the top. The solid green line represents the best fit relation reported by \citet{Bianchi2007}. The combined error range on the slope and normalization of the best fit is also shown by dotted lines in the respective colors.}
\label{fig: Lxvseqw-fit}
\end{figure*}

We then investigated for a dependence of the Fe K$\alpha$ line strength on the X-ray luminosity. Figure \ref{fig: Lxvseqw-fit} shows the relation between the average rest frame EW and 2-10 keV luminosity in all redshift and their associated luminosity bins. The weakening of the line strength with increasing luminosity is clearly present. To derive the relation between the two parameters we followed the method of \citep{Guainazzi2006}. Their method can be summarized as follows. First, a set of Monte-Carlo simulated data is derived from the EWs (both upper limits and measurements with their associated symmetric errors) according to the following rules: (i) each measured EW was substituted by a random Gaussian distribution, whose mean is the best fit value and whose standard deviation is its statistical uncertainty; (ii) each upper limit (U) was substituted by a random uniform distribution in the interval [0,U]. Then, ordinary least squares fits were performed on each Monte-Carlo simulated data set. The mean of the slopes and the intercepts derived from the fits of each data set are taken as the ``best fit parameters of the linear relation''. We obtained the following best fit relation:


\begin{equation} \label{ITeqn}
\log(EW_{Fe}) = (1.66\pm0.09)+(-0.43\pm0.07) \log(L_{X,44})
\end{equation}

\noindent where EW$_{Fe}$ is the equivalent width of the narrow Fe K$\alpha$ line in eV and L$_{X,44}$ is the 2-10 keV X-ray luminosity in units of 10$^{44}$ erg s$^{-1}$. To test the significance of this anti-correlation, we calculated the Spearman's rank coefficient (R$_S$) for each Monte Carlo simulated data set. The mean value is taken as the R$_S$ of Equation \ref{ITeqn}. The R$_S$ for this correlation is -0.74, corresponding to a Null Hypothesis Probability (NHP) of 2.95$\times$10$^{-4}$.

Our data therefore strongly imply that the Fe K$\alpha$ line strength in the selected samples decreases as a function of increasing luminosity in all redshift bins confirming the IT effect. In the previous Subsection \ref{subsec-line-z-dependence}, we explored the redshift dependence of the Fe K$\alpha$ line intensity. The observed non evolutionary behavior assured us that no additional redshift effects are introduced in the IT relation given by Equation \ref{ITeqn}, which includes different luminosities and therefore on average different redshifts. This indicates that the driver of the observed IT correlation is luminosity and not redshift. 

However, the slope of the EW$_{Fe}$ versus L$_X$ relation we obtained is steeper than the value reported in previous studies (-0.20$\pm$0.03 \citep{IT1993}; -0.17$\pm$0.08 \citep{Page2004}; -0.17$\pm$0.03 \citep{Bianchi2007}). The normalization of our EW$_{Fe}$ versus L$_X$ relation is statistically consistent with that reported by \citet{Bianchi2007}. To check whether this discrepancy in the slope is due to the 2 extreme points at L$_X$$<$10$^{42}$ erg s$^{-1}$, we removed these points and subsequently performed the linear fits. We still obtained a slope of -0.43$\pm$0.09 with R$_S$=-0.67 and NHP=3.24$\times$10$^{-3}$. 

\begin{figure}
\begin{center}
\begin{tabular}{c}
\resizebox{80mm}{!}{\includegraphics[angle=0]{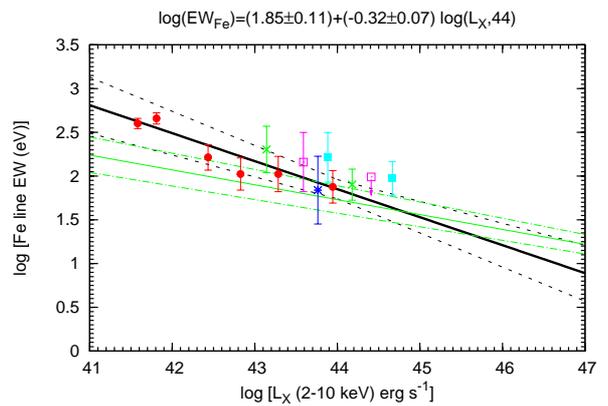}}
\end{tabular}
\caption{Rest frame equivalent width of the narrow Fe K$\alpha$ line in the redshift interval 0$<$z$<$0.8 plotted against the X-ray luminosity in the 2-10 keV band. (Symbols, line styles and colors same as in Figure \ref{fig: Lxvseqw-fit}).}
\label{fig LX-EW-fit-subset}
\end{center}
\end{figure}

This discrepancy in the slope can possibly be attributed to the difference in composition of the samples. Indeed, the sample in \citet{Bianchi2007} consists of 157 local unobscured radio quiet AGNs (almost 90\% within z$<$1), whereas our sample comprises a mixture of 507 obscured and unobscured sources over a broader z range (69\% at z$<$1 and 31\% at z$>$1). The difference in the slope we obtain for the whole sample can also be driven by the upper limits in the highest redshift bins which are not sampled by \citet{Bianchi2007}. When all bins in the redshift range 0$<$z$<$0.8 are considered (see Figure \ref{fig LX-EW-fit-subset}), the inverse correlation between the EW and X-ray luminosity becomes 

\begin{equation}\label{eqn2}
 \log(EW_{Fe}) = (1.85\pm0.11)+(-0.32\pm0.07) \log(L_{X,44})
\end{equation}

\noindent with R$_S$=-0.64 and NHP=3.63$\times$10$^{-2}$. The slope of the anticorrelation is statistically consistent in Equation \ref{ITeqn} and \ref{eqn2}.
 However, the best fit value is flatter in Equation \ref{eqn2}. This in turn supports the hypothesis that upper limits are an important factor in the calculation of the fit relation. 

The physical origin of the IT effect is related to where the line is originated. Several studies have confirmed that the narrow neutral Fe K$\alpha$ line is a ubiquitous component in AGNs spectra \citep{Page2004, Yaqoob2004, Corral2008}. The measured widths vary substantially from source to source and cover a wide range from 1000 to 15,000 km s$^{-1}$ \citep{Yaqoob2004}, thus being consistent either with an origin from the broad line region (BLR) or the torus (obscuring circumnuclear matter) envisaged in orientation-dependent unification scheme for AGNs \citep{Urry1995}. The original IT effect \citep{IT1993} was attributed to a decrease in the covering factor of the BLR with increasing X-ray luminosity. However, recent studies report that no strong correlation exists between the Fe K$\alpha$ core width and prominent BLR H$\beta$ $\lambda$4861 line width \citep{Nandra2006}. There is a similar lack of correlation between the EWs of the Fe K$\alpha$ and CIV $\lambda$1549 lines \citep{Wu2009}. This implies that the Fe K$\alpha$ line is unlikely to be produced in the BLR. 

Based on previous studies, the abundance effect and black hole mass can be ruled out as the primary physical parameters driving the correlation \citep{Page2004, Zhou2005, Nandra2006, Bianchi2007}. The absence of evolution of the Fe K$\alpha$ EW with redshift (see Subsection \ref{subsec-line-z-dependence}) further confirms this hypothesis. Assuming that the line is originated in the torus, the luminosity dependent covering factor of the torus turns out to be the most likely explanation for the IT effect \citep{Page2004, Bianchi2007}. A lower covering factor signifies that the torus obscures a smaller solid angle around the nucleus, and hence less Fe K-edge photons are captured and generate less Fe K$\alpha$ photons, leading to a reduction in the EW. In this way, an increase in luminosity and a subsequent decrease in the covering factor could explain the observed anticorrelation of the EW and X-ray luminosity. 

\begin{figure}
\centering
\includegraphics[width=7cm]{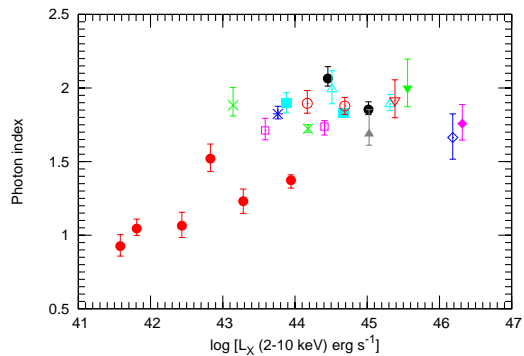}
\caption{The variation of the powerlaw photon index as a function of the rest frame 2-10 keV luminosity is illustrated. Symbols reflect different redshift bins as in Figure \ref{fig: Lxvseqw-fit}.}
\label{fig: zvar-nh}
\end{figure}

An immediate implication of the luminosity dependent covering factor of the torus is the variation in the underlying continuum with luminosity as high covering factor at low luminosities would cause spectral hardening by suppressing the soft part of the spectrum. To confirm this argument, we investigated the dependence of the power law photon index (characterizing the underlying continuum) on the X-ray luminosity in different redshift bins. In order to avoid the complexity of the AGNs spectra at energies below 2 keV, we performed spectral analysis in the 2-10 keV rest frame band. Figure \ref{fig: zvar-nh} illustrates the variation of the power law photon index ($\Gamma$) with the 2-10 keV luminosity in different redshift intervals. We found that the $\Gamma$ in our stacked spectra sample is hard at low X-ray luminosities (L$_X$$<$10$^{43}$ erg s$^{-1}$) sampled mostly in the redshift bin 0.0-0.2 (see also Figure 2). A hard $\Gamma$ is indeed expected in the case of increasing absorption at these luminosities due to a high covering factor. The trend also reflects the decrease of the fraction of the absorbed sources with the X-ray luminosity revealed from extensive works in hard X-ray surveys studies \citep{Ueda2003, LaFranca2005, Hasinger2008}.
 
\section{Conclusions}
We have compiled a large sample of 507 sources selected from the 2XMM catalogue covering the redshift range 0$<$z$<$5 and nearly 6 orders of magnitude in the X-ray luminosity. To improve the signal-to-noise ratio and investigate the average X-ray spectral properties of the sources, we computed the integrated spectra in narrow intervals of redshift and luminosity. We performed simulations to assess the accuracy of the stacking procedure. The main results can be summarized as follows.
\begin{itemize}
\item[$\bullet$] {We confirm that the narrow Fe K$\alpha$ line is an almost ubiquitous feature in the X-ray spectra of AGNs.}
\item[$\bullet$] {In the integrated spectra spanning redshift range 0$<$z$<$1.3 at nearly constant luminosity we detect no strong trend in the Fe K$\alpha$ line EW with redshift.}
\item[$\bullet$]{We find a significant anticorrelation between the Fe K$\alpha$ EW and 2-10 keV rest frame luminosity across a wide redshift range not probed in prior studies. The slope of the anticorrelation we obtained is steeper than the value reported in previous studies, likely due to the difference in composition of the samples. The exact physical cause of the IT effect is unknown, one promising explanation is a decline in the covering factor of the putative molecular torus as the luminosity increases.}
\item[$\bullet$]{We find a hardening of the spectral indices at low luminosities in the redshift bin 0.0-0.2. This can be linked to the dependence of obscuration with luminosity.}
\end{itemize}

Large area X-ray surveys from sensitive next generation satellites (e.g., eROSITA and WFXT), when coupled with deep multiwavelength photometry (e.g., PanSTARRS and LSST) and spectroscopic campaigns (SDSS, LAMOST, and Euclid) will allow to study the dependence of the EW strength with redshift and luminosities on samples of several order of magnitudes larger number statistics providing a unique sampling of the L-z plane.

\acknowledgement
Our results are based on observations obtained with XMM-{\it Newton}, an ESA science mission with instruments and contributions directly funded by ESA member states and the USA (NASA). The XMM-{\it Newton} project is supported by the Bundesministerium f\"ur Wirtschaft und Technologie/Deutsches Zentrum f\"ur Luft und Raumfahrt (BMWI/DLR, FKZ 50 OX 0001), the Max-Planck Society and the Heidenhain-Stiftung and also by PPARC, CEA, CNES, and ASI. This research has made use of the NASA/IPAC Extragalactic Database (NED) which is operated by the Jet Propulsion Laboratory, California Institute of Technology, under contract with the National Aeronautics and Space Administration. GH acknowledges support from the German Deutsche Forschungsgemeinschaft, DFG, Leibniz Prize (FKZ HA 1850/28–1). This work is partially supported by ASI-INAF grants I/023/05/00 and I/088/06. This research was also supported by the DFG cluster of excellence ``Origin and Structure of the Universe". The authors would like to thank the anonymous referee for the encouraging report and useful comments, which improved the paper significantly. We also thank Stefano Bianchi and Kazushi Iwasawa for useful discussions.

\bibliographystyle{aa}
\bibliography{references}

\end{document}